\documentclass[aps,pra,showpacs,amsmath,amssymb,superscriptaddress,reprint,10pt,longbibliography]{revtex4-1}
\usepackage{graphicx,mathrsfs,enumerate}
\usepackage[utf8]{inputenc}
\usepackage{mathtools}
\usepackage{amsfonts,amstext}
\usepackage{amsmath,amsthm,amssymb,mathbbol,tensor}
\usepackage{algorithmicx}
\usepackage{graphicx,xcolor,mathrsfs,enumerate}
\usepackage[hidelinks]{hyperref}\hypersetup{pdfpagemode=UseNone,pdfstartview=}
\usepackage[left=2cm,right=2cm]{geometry}
\usepackage{url}
\usepackage{times}
\usepackage{soul}
\usepackage{bbm}
\usepackage[english]{babel}

\DeclareMathOperator{\Tr}{Tr}

\newtheorem{dfn}{Definition}

\newtheorem{thm}{Theorem}

\newtheorem{prot}{Protocol}

\newcommand{\asmb}[1]{\boldsymbol{#1}_{A|X}}

\newcommand{\asmbx}[2]{\boldsymbol{#1}^{#2}_{A|X}}
\newcommand{\afid}{\mathcal{F}_{\textrm{A}}}
\newcommand{\singfid}{\mathcal{F}_{\Phi}}

\begin{document}
\title{Distillation of quantum steering}
\author{R. V. Nery}
\affiliation{Instituto de F\'isica, Universidade Federal do Rio de Janeiro, P. O. Box 68528, Rio de Janeiro, RJ 21941-972, Brazil}
\author{M. M. Taddei}
\affiliation{Instituto de F\'isica, Universidade Federal do Rio de Janeiro, P. O. Box 68528, Rio de Janeiro, RJ 21941-972, Brazil}
\author{P. Sahium}
\affiliation{Instituto de F\'isica, Universidade Federal do Rio de Janeiro, P. O. Box 68528, Rio de Janeiro, RJ 21941-972, Brazil}
\author{S. P. Walborn}
\affiliation{Instituto de F\'isica, Universidade Federal do Rio de Janeiro, P. O. Box 68528, Rio de Janeiro, RJ 21941-972, Brazil}
\author{L. Aolita}
\affiliation{Instituto de F\'isica, Universidade Federal do Rio de Janeiro, P. O. Box 68528, Rio de Janeiro, RJ 21941-972, Brazil}
\author{G. H. Aguilar}
\affiliation{Instituto de F\'isica, Universidade Federal do Rio de Janeiro, P. O. Box 68528, Rio de Janeiro, RJ 21941-972, Brazil}

\begin{abstract}
We show -- both theoretically and experimentally -- that Einstein-Podolsky-Rosen steering can be distilled. We present a distillation protocol that outputs a perfectly correlated system -- the singlet assemblage -- in the asymptotic  infinite-copy limit, even for inputs that are arbitrarily close to being unsteerable. As figures of merit for the protocol's performance, we introduce the assemblage fidelity and the singlet-assemblage fraction. These are potentially interesting quantities on their own beyond the current scope. Remarkably, the protocol works well also in the non-asymptotic regime of few copies, in the sense of increasing the singlet-assemblage fraction. We demonstrate the efficacy of the protocol using a hyperentangled photon pair encoding two copies of a two-qubit state. This represents to our knowledge the first observation of deterministic steering concentration. Our findings are not only fundamentally important but may also be useful for  semi device-independent protocols in noisy quantum networks.
\end{abstract}



\maketitle


Steering is a unique form of quantum nonlocality that appears in hybrid quantum networks with both trusted and untrusted components \cite{WJD2007}. The first prototypes of the quantum internet may implement such a scenario, where only few members of the network (such as big servers) would have the resources to fully characterize, and therefore trust, their devices, while the remaining participants (such as the end-users) would operate the untrusted devices. These scenarios are referred to as semi device-independent (DI), in contrast to the fully DI context, where all apparatuses are untrusted, or the device-dependent one, with trusted components exclusively. A trusted device allows for full quantum control of the system it operates, e.g. through well-characterized quantum measurements on it. A device is untrusted if one can only control its classical settings (inputs), obtaining classical outcomes (outputs) of uncharacterised measurements from it, thus effectively working as a black-box device. Importantly, steering certifies the presence of entanglement in a semi-DI fashion. 
Due to this, apart from its fundamental relevance, it is  important also from an applied point of view: Steering is known to be the key resource behind several information-processing tasks in the semi-DI scenario \cite{BCWSW12, PW15}.

However, as experimental quantum networks grow ever more complex, the unavoidable noise and imperfections become increasingly significant. This can severely degrade the steering in the network, compromising the performance of the implemented task. Distillation protocols are ideal for these situations, as they concentrate the resource contained in multiple copies of a noisy system into a pure maximally-resourceful system, which can then be directly used safely for the task in question. Interestingly, distillation protocols are known for the other two paradigmatic variants of quantum nonlocality--namely, entanglement in the device-dependent framework and Bell nonlocality in the fully DI one--\cite{BennettDist_PRA1996,Forster2009} and also even for other important quantum resources \cite{Kitaev2005, Spekkens2008, Brandao2013, Winter2016, Morris_WorkDist2019, Taddei2019}. Nevertheless, to our knowledge, almost nothing is known for the case of steering. The related phenomenon of steering superactivation \cite{Quintino2016, Quintino_Activ2016, Pramanik2019} is known to exist, but requires complete quantum control on all parties. In particular, it is not known whether steering distillation exists (in the asymptotic regime of infinitely many copies of the noisy system) or even if steering can be partially purified in the finite-copy regime.

Here, we answer both questions in the affirmative. We theoretically prove that steering distillation exists, devising an explicit simple protocol for it. We show that such protocol not only distills pure \emph{singlet assemblages} (i.e., the steering correlations generated by a maximally-entangled singlet state under ideal von Neumann measurements) in the asymptotic infinite-copy regime but it also succeeds at concentrating steering in the finite-copy regime. Moreover, the proposed protocol distills quantum steering in a hybrid scenario, where only one device applies controlled operations, while the other just deals with its inputs and outputs. This is conceptually different from entanglement distillation, where fully-quantum local operations and classical communication (LOCC) are applied on both sides. Remarkably, we demonstrated that an initial system with an arbitrarily small (constant) amount of pure steering can be distilled. To quantify the performance of the partial purification in the finite-copy regime, we introduce the \emph{assemblage fidelity} as a measure of closeness between the steering correlations of two different systems. Finally, we experimentally demonstrate the efficacy of the protocol in an optical setup with 2 hyperentangled photons, encoding 2 copies of a 2-qubit state each. We observe a clear increase of the protocol's output's \emph{singlet-assemblage fraction} (the assemblage fidelity with respect to the singlet assemblage).

\begin{figure}
\centering
\includegraphics[width=0.95\linewidth]{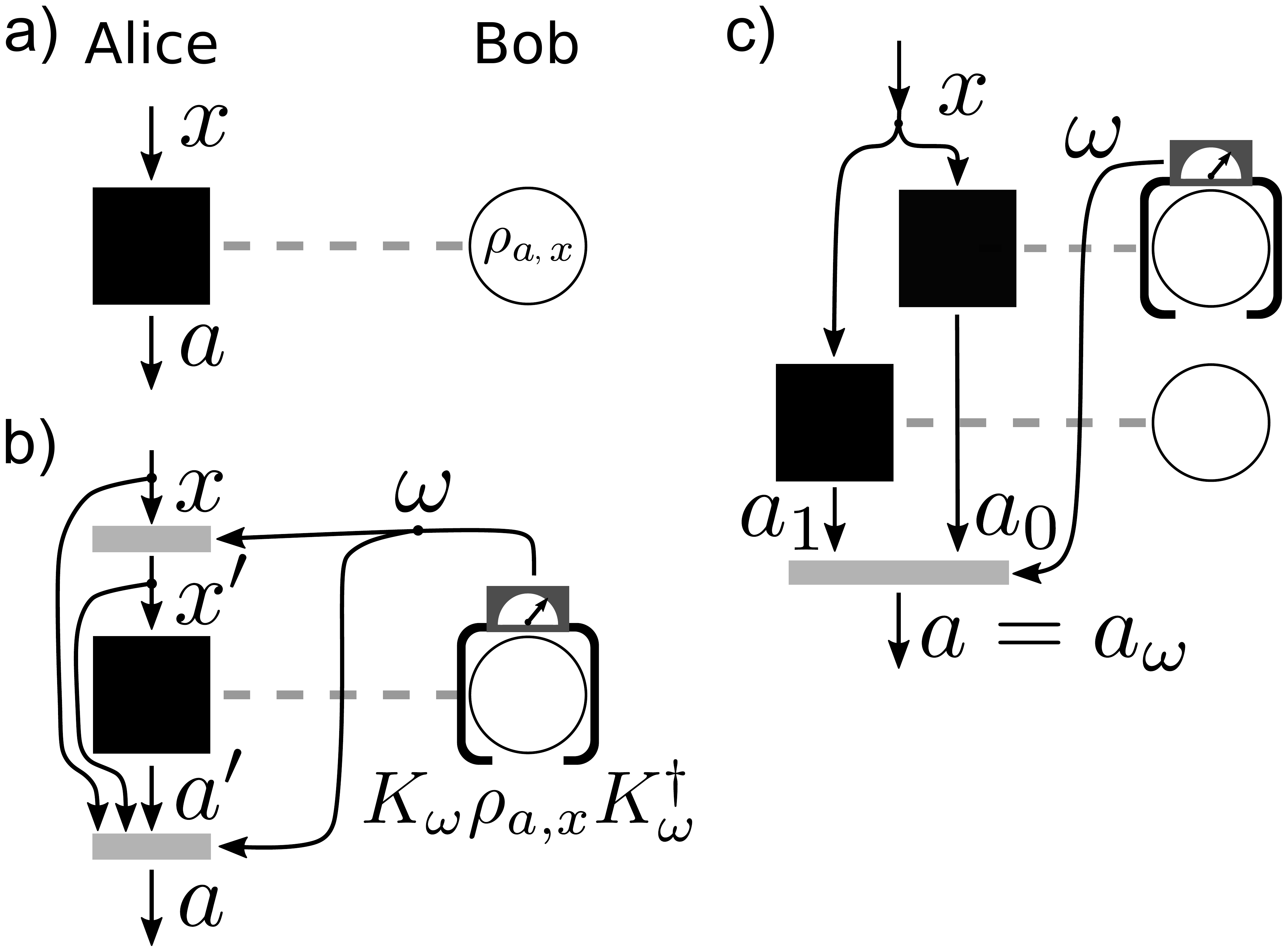}
\caption{a) Bipartite semi-DI scenario. Alice can only perform uncharacterized measurements on her device, which is then effectively treated as a black box. Bob has full quantum control on his system, allowing complete knowledge of his quantum state. Together with Alice's probability distribution for her black box, the systems compose the assemblage $\asmb{\Sigma}$. b) Depiction of a generic 1WLOCC. The assemblage can be manipulated locally by both parties and Bob is allowed to communicate any classical parameter to Alice. Alice's wirings, represented by gray boxes, allow creation of new random variables from previous ones by an arbitrary distribution. c) Protocol \ref{protocol} for two copies of the original assemblage. Bob applies a local filter on one of his qubits. A successful outcome results in a singlet assemblage shared between Alice and Bob, while failure produces an unsteerable assemblage. In both cases, communication of Bob's result to Alice is used and the appropriate subsystem is then chosen by both parties.
}
\label{Fig:StDist_1WLOCC}
\end{figure}

\emph{Preliminary definitions.}-- We consider two parties, Alice and Bob, sharing initially a correlated system in a semi-DI scenario [ Fig.\ \ref{Fig:StDist_1WLOCC} (a)]. We assume that Alice has an untrusted black-box device, while Bob  holds a fully-characterized trusted quantum device. Alice's input is represented by a classical parameter $x\,\in\,[m]$, where $[n]$ is a shorthand notation for $\{0,1,...,n-1\}$ for any natural number $n$, and $m \geq 2$ is the number of possible settings. For a given choice $x$, a classical output $a \in [o]$ is obtained from the black box, where $o \geq 2$ is the number of possible outputs. Complete characterization of Bob's device allows him to reconstruct his quantum state.

The system is then completely described by the conditional distribution $\{P(a|x)\}_{a \in [o],x \in [m]}$ of obtaining an output $a$ from the black box given the input choice $x$, and by the conditional quantum state $\rho_{a,x}$ on Bob's side, supported on a local Hilbert space $\mathcal{H}_B$ and possibly correlated to Alice's variables. Equivalently, both objects can be neatly encapsulated in the so-called \emph{assemblage} \cite{WJD2007}, which is the list $\boldsymbol{\Sigma}_{A|X} \coloneqq~\{ \sigma_{a|x}\}_{a \in [o], x \in [m]}$ of subnormalized bounded operators $\sigma_{a|x}$ supported also on $\mathcal{H}_B$, such that $ \Tr[\sigma_{a|x}] = P(a|x)$ and $\rho_{a,x} = \sigma_{a|x}/\Tr[\sigma_{a|x}]$.

The assemblage is said to be \emph{steerable} if it cannot be written in terms of a local-hidden-states (LHS) model. This means that there exists no hidden variable that turns Bob's state statistically independent from Alice's variables; i.e., an assemblage $\asmbx{\Sigma}{\textrm{LHS}}\coloneqq \{\sigma^{\mathsf{LHS}}_{a|x}\}_{a,x}$ admits an LHS model if there is a variable $\Lambda$ admitting values $\lambda \in [L]$, such that
\begin{equation}
\sigma^{\mathsf{LHS}}_{a|x} = \sum_\lambda P_\Lambda(\lambda)\,P(a|x,\lambda)\,\rho_\lambda.
\label{eq:LHS_def}
\end{equation}
$L$ is the number of possible configurations for $\Lambda$, and $P_\Lambda$ is their probability distribution. Bob's local hidden states $\rho_\lambda$ are independent of Alice's variables $a$ and $x$ in this case.

In the semi-DI scenario, operations are restricted due to the lack of characterization of Alice's device. Free operations--i.e. operations that do not create quantum steering out of LHS assemblages--are restricted to 
1-way local operations and classical communication (1WLOCCs) \cite{Gallego2015}. These correspond to local pre- and post-processing operations of Alice's classical inputs and outputs, respectively, conditioned on the outputs of quantum operations on Bob's side. Examples of these are shown in Fig.\ \ref{Fig:StDist_1WLOCC}. 


\emph{Distillation of quantum steering.}-- The task of steering distillation consists of extracting from $N$ copies of a weakly steerable assemblage a smaller number of an extremal assemblage with a purer form of steering, using free operations only. We consider as target here a singlet assemblage, i.e. an assemblage obtained from a singlet state by orthogonal rank-1 projective measurements on Alice's side. These are extremal in the sense that they cannot be obtained from other singlet assemblages via 1WLOCC \cite{Gallego2015}. Furthermore, they are known to maximize important measures of quantum steering \cite{PW15, SABGS16, Kaur2017}. In particular, we consider the singlet assemblage $\asmbx{\Sigma}{\Phi^+}$ obtained from the maximally entangled state $|\Phi^+\rangle \coloneqq (|00\rangle + |11\rangle)/\sqrt{2}$ when Alice's measurements correspond to the Pauli matrices $Z$ and $X$. This assemblage is characterized by the components
\begin{subequations}
\label{eq:SingAsmb}
\begin{align}
\sigma^{\Phi^+}_{0|0} = \frac12 |0\rangle\langle 0|,&\quad 
\sigma^{\Phi^+}_{1|0} = \frac12 |1\rangle\langle 1|, 
\label{eq:SingAsmb_x0}\\
\sigma^{\Phi^+}_{0|1} = \frac12 |+\rangle\langle +|,&\quad 
\sigma^{\Phi^+}_{1|1} = \frac12 |-\rangle\langle -|,
\label{eq:SingAsmb_x1}
\end{align}
\end{subequations}
where $|\pm\rangle \coloneqq (|0\rangle \pm |1\rangle)/\sqrt{2}$.

Then, we can now define the task of steering distillation (with respect to $\asmbx{\Sigma}{\Phi^+}$ as target assemblage) as the following assemblage conversion
\begin{equation}
(\asmb{\Sigma})^{\otimes N}\,\xrightarrow{\textrm{1WLOCC}}\,(\asmbx{\Sigma}{\Phi^+})^{\otimes r N},
\label{eq:DistTask}
\end{equation}
with unit probability as $N \rightarrow \infty$ and with $0 < r \leq 1$ the distillation rate of the protocol. The initial resource of the process is given by $N$ independent copies of the assemblage $\asmb{\Sigma}$, which is represented mathematically as $\asmb{\Sigma}^{\otimes N} \coloneqq \{ \otimes_{i=1}^N\,\sigma_{a_i|x_i}\}_{a_1,x_1,...,a_N,x_N}$.


In what follows, we will assume that Alice and Bob share initially $N \geq 2$ copies of the pure non-orthogonal assemblage $\asmbx{\Sigma}{(\alpha)} \coloneqq \{ \sigma^{(\alpha)}_{a|x}\}_{a,x}$, obtained from the state
\begin{equation}
|\alpha\rangle \coloneqq \alpha|00\rangle + \beta|11\rangle,
\label{eq:StatePhiAlpha}
\end{equation}
through $Z$ and $X$ Pauli measurements on Alice's side, where  $0<\beta< \alpha<1$ and $\alpha^2+\beta^2=1$. The assemblage is then characterized by the components
\begin{subequations}
\label{AsmbAlpha}
\begin{align}
\sigma^{(\alpha)}_{0|0} = \alpha^2\,|0\rangle\langle 0|,&\quad
\sigma^{(\alpha)}_{1|0} = \beta^2\,|1\rangle\langle 1|,
\label{AsmbAlpha_Comp10} \\
\sigma^{(\alpha)}_{0|1} = \frac{1}{2}\,|\alpha_+\rangle\langle \alpha_+|,
&\quad \sigma^{(\alpha)}_{1|1} = \frac{1}{2}\,|\alpha_-\rangle\langle \alpha_-|,
\label{AsmbAlpha_Comp11}
\end{align}
\end{subequations}
where $|\alpha_{\pm}\rangle \coloneqq \alpha|0\rangle \pm \beta|1\rangle$.

We also consider a dichotomic POVM $\boldsymbol{M} \coloneqq \{ M^{(0)}, M^{(1)}\}$ on Bob's subsystem, where $M^{(\omega)}$ are bounded operators satisfying $M^{(\omega)} \geq 0$ and $M^{(0)} + M^{(1)} = \mathbbm{1}$. We say that $\boldsymbol{M}$ is applied on an assemblage $\asmb{\Sigma}$ when Bob applies the corresponding POVM on his quantum state. When outcome $\omega$ is obtained, the assemblage's components are updated by \cite{Gallego2015, Note1}
\begin{equation}
\sigma^\prime_{a|x,\omega} = \frac{\sqrt{M^{(\omega)}}\,\sigma_{a|x}\,\sqrt{M^{(\omega)}}^{\,\dagger}}{\Tr\left[M^{(\omega)}\,\rho_B\right]},
\label{eq:AsmbPOVM}
\end{equation}
where $\rho_B \coloneqq \sum_a \sigma_{a|x}$ is Bob's reduced state (well-defined by virtue of the no-signalling principle \cite{Brunner2014, Cavalcanti2016}). Introducing the notation $K^{(\omega)} \coloneqq \sqrt{M^{(\omega)}}$, so that $M^{(\omega)} = K^{(\omega)\,\dagger} K^{(\omega)}$, we can now present our protocol:
\begin{prot}[Local filtering with one-sided quantum control]
\label{protocol}
Let Alice and Bob share $\asmbx{\Sigma}{(\alpha)\,\otimes N}$, with $N \geq 2$, and let  $\boldsymbol{M}$ be a dichotomic POVM of elements
\begin{subequations}
\label{eq:KrausRep}
\begin{align}
K^{(0)} &\coloneqq \frac{\beta}{\alpha}\,|0\rangle\langle 0| + |1\rangle\langle 1|,
\label{eq:KrausRep_OutSuccess}
\\
K^{(1)} &\coloneqq \frac{\sqrt{\alpha^2 - \beta^2}}{\alpha}\,|0\rangle\langle 0|.
\label{eq:KrausRep_OutFail}
\end{align}
\end{subequations}
Then, 
\begin{algorithmic}[1]

\State For $1\leq i\leq N-1$, Bob measures $\boldsymbol{M}$ on each $i$-th copy of $\asmbx{\Sigma}{(\alpha)}$ and gets an outcome $\omega_i\in\{0,1\}$.

\State If $\omega_i=1$ for all $1\leq i\leq N-1$, he sets $\omega_N=0$ without measuring the last copy; otherwise he sets $\omega_N=1$. Then, he sends the string $\boldsymbol{\omega}:=\omega_1,\hdots \omega_N$ to Alice.

\State Alice gets $\boldsymbol{\omega}$. Then, Alice and Bob discard every $i$-th system for which $\omega_i=1$, for all $1\leq i\leq N$. The output of the protocol is given by the remaining  assemblages.
\end{algorithmic}
\end{prot}
\noindent The protocol is depicted in Fig.\ \ref{Fig:StDist_1WLOCC} (c)  for $N=2$ copies of the initial assemblage.


Any steering distillation protocol must guarantee extraction of at least one such singlet assemblage in the regime of asymptotically many copies, $N\,\rightarrow\,\infty$. For a finite number of copies, however, perfect extraction may not be possible and only an approximation of $\asmbx{\Sigma}{\Phi^+}$ is attainable. To quantify this notion of proximity and have a figure of merit for the protocol for finite $N$, we define the following quantities:
\begin{dfn}[Assemblage fidelity]
Let $\asmb{\Sigma} = \{ \sigma_{a|x}\}_{a \in [o],x \in [m]}$ and $\asmb{\Xi} = \{ \xi_{a|x}\}_{a \in [o],x \in [m]}$ have the same number of inputs and outputs and act on the same Hilbert space $\mathcal{H}_B$. We define the \emph{assemblage fidelity} between $\asmb{\Sigma}$ and $\asmb{\Xi}$ as
\begin{equation}
\afid(\asmb{\Sigma},\,\asmb{\Xi})\coloneqq \min_{x\in[m]} \sum_{a \in [o]}\,\mathcal{F}(\sigma_{a|x},\xi_{a|x}),
\label{eq:Asmb_Fidelity}
\end{equation}
with $\mathcal{F}(A,B) =~\Tr\left[\sqrt{\sqrt{A}B\sqrt{A}}\right]$ the usual state fidelity between two density matrices $A$ and $B$ on $\mathcal{H}_B$. 
\end{dfn}
The definition of assemblage fidelity retains many of the expected properties for a fidelity-like quantity from its dependence on the usual fidelity $\mathcal{F}$, see the Supplemental Material (SM) for demonstrations \cite{SuppMat}.
In particular, $\afid$ is nonnegative and $\afid(\asmb{\Sigma}, \asmb{\Xi}) \leq 1$, with equality holding iff $\asmb{\Sigma} = \asmb{\Xi}$. It should be remarked that the minimization contained in definition \eqref{eq:Asmb_Fidelity} is used precisely to preserve these properties and should be understood as a way of better distinguishing assemblages that are in fact distinct. See the SM for details \cite{SuppMat}.

Assume now that the assemblage $\asmb{\Sigma}$ is defined in the same space of $\asmbx{\Sigma}{\Phi^+}$, defined in Eqs. \eqref{eq:SingAsmb}. Then, we can now define our figure of merit for the distillation protocol's performance:
\begin{dfn}[Singlet-assemblage fraction]
Let $\asmbx{\Sigma}{\Phi^+}$ be the singlet assemblage defined in Eqs. \eqref{eq:SingAsmb}. We define the \emph{singlet-assemblage fraction} for an assemblage $\asmb{\Sigma} = \{\sigma_{a|x}\}_{a \in [2],\,x \in [2]}$, with $\mathrm{dim}(\mathcal{H}_B) = 2$, as  
\begin{equation}
\singfid(\asmb{\Sigma}) \coloneqq \afid(\asmb{\Sigma},\asmbx{\Sigma}{\Phi^+}).
\label{eq:SingFid_def}
\end{equation}
\end{dfn}
With this quantity, we may evaluate if a given protocol indeed allows extraction of an assemblage that is closer to the singlet assemblage than initially. Ideally, the singlet-assemblage fraction should be defined including an optimization over unitaries applied on Bob's side (or, more generally, over reversible 1WLOCCs). This however enormously complicates its analytical computation even for the case considered in our results below, where we observe numerically that the values with and without this extra optimization coincide. We now present our main result, proven in App. B in the SM \cite{SuppMat}.
\begin{thm}[Distillation of Quantum Steering]
\label{Thm:SteerDistillation}
Quantum steering can be distilled with the use of protocol \ref{protocol} with rate $r=2\,\beta^2$ in the asymptotic regime of infinite copies of the initial assemblage $\asmbx{\Sigma}{(\alpha)}$. 
Furthermore, in the regime of $N$ copies, with $N$ finite, an assemblage can be obtained on average which is closer to the singlet assemblage than $\asmbx{\Sigma}{(\alpha)}$, attaining a singlet-assemblage fraction of $\sqrt{1 - \frac12\,(\alpha - \beta)^2(\alpha^2 - \beta^2)^{N-1}}$.
\end{thm}

\begin{figure}[t]
\centering
\includegraphics[width = 0.85\linewidth]{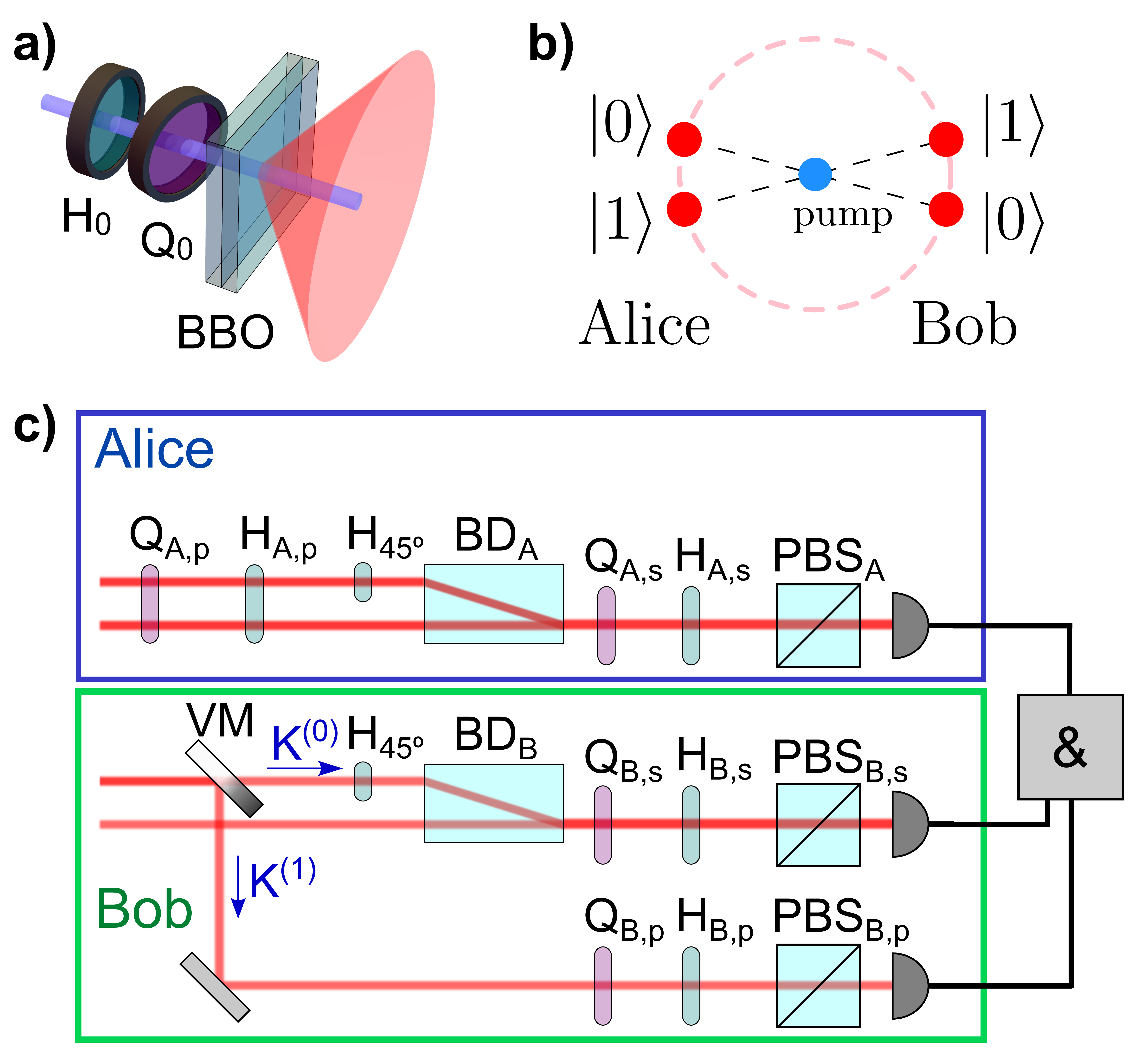}
\caption{a-b) Production of entangled photons via SPDC. The down-converted light is spectrally filtered to $(650\,\pm \,10)$ nm and collimated by a lens (not shown in figure), which converts the photons' momentum to spatial modes parallel to the pump beam. Only two pairs of correlated spatial modes are used in the remainder of the setup, as shown in (b), corresponding to two additional qubits besides polarization. c) Setup for quantum steering distillation. Alice's interferometer allows for measurements both on polarization and on the spatial DOF and comprises Alice's two initial black boxes. On Bob's side, the amplitude filter is implemented by the variable reflectivity mirror (VM); reflectivity is tuned so that spatial DOF amplitudes become equalized when the photon is transmitted. Polarization is ignored and tomography of spatial mode DOF ensues in the upper branch of Bob's setup. 
Legend for the components: Q - Quarter-waveplate; H - Half-waveplate; BD - Beam displacer; PBS - Polarizing beam splitter; VM - Variable reflectivity mirror. Subindices indicate to which party (A - Alice, B - Bob) and to which type of measurement (p - polarization, s - spatial mode) the component pertains. }
\label{Fig:StDist_Setup}
\end{figure}

\emph{Experimental realization.}-- We implemented the local filtering protocol experimentally using two copies of the original assemblage. A pair of hyperentangled photons in polarization and optical path, produced via spontaneous parametric downconversion (SPDC), is used to encode the two copies, one in each degree of freedom (DOF). The setup is represented in Fig.\ \ref{Fig:StDist_Setup}. A 325-nm continuous-wave He-Cd laser pumps two type-I beta-barium borate (BBO) crystals in a cross-axis configuration \cite{Kwiat99}, generating photon pairs centered at 650 nm. Waveplates $H_0$ (half) and $Q_0$ (quarter) are set to produce photons in a polarization state close to $|\alpha\rangle$ [Eq.\ \eqref{eq:StatePhiAlpha}]. We use the encoding $|H\rangle \rightarrow |0\rangle,\,|V\rangle \rightarrow |1\rangle$, where $|H\rangle$ and $|V\rangle$ correspond to horizontal and vertical polarizations, respectively.  Different values of $\alpha$ are realized by varying the angle on $H_0$. By keeping only two correlated directions produced in the SPDC we define the optical path qubits [Fig.\ \ref{Fig:StDist_Setup} (b)]. Path-dependent attenuators are used to make amplitudes match those of $|\alpha\rangle$. With this, we obtain another copy of the initial state between the parties.

\begin{figure}
\centering
\includegraphics[width = 0.95\linewidth]{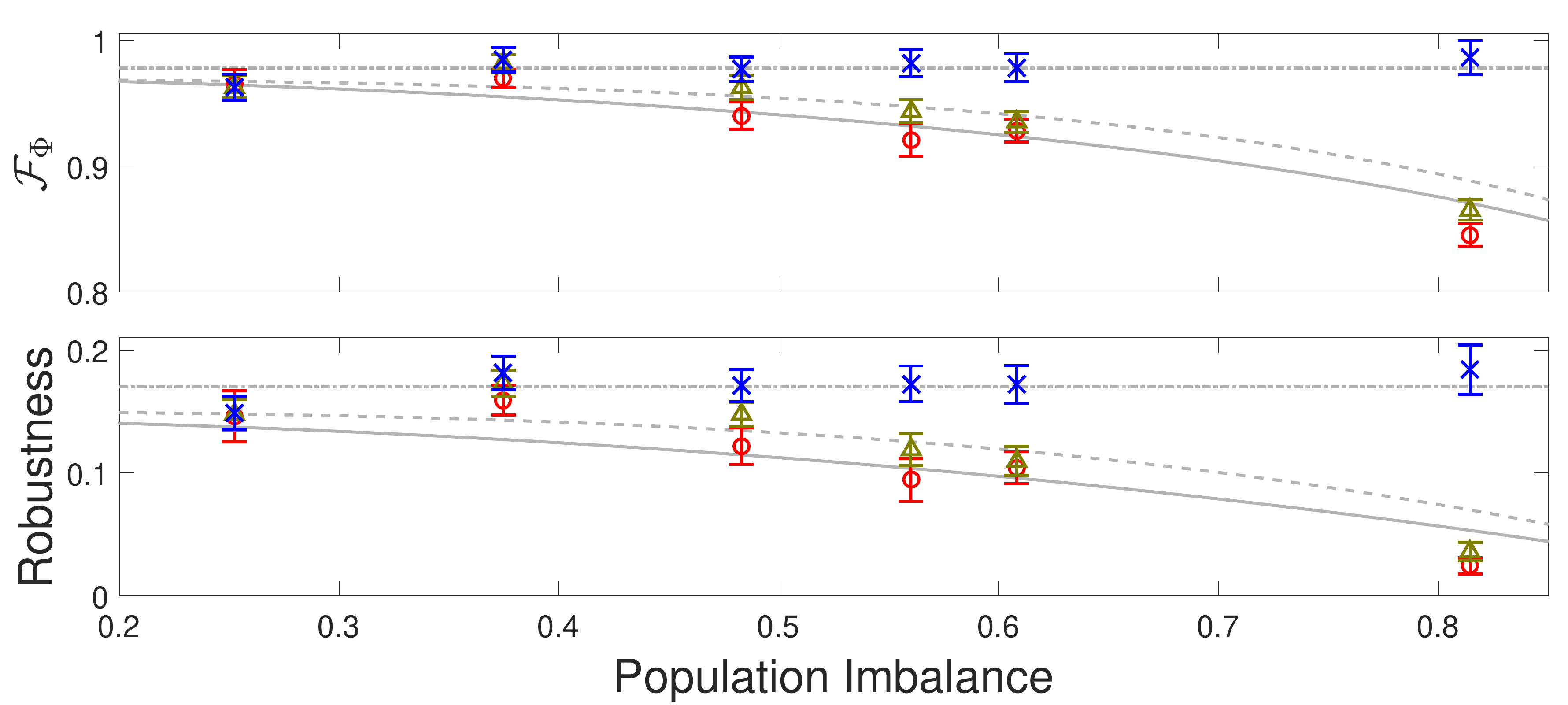}
\caption{Singlet-assemblage fraction (top) and LHS-robustness of steering (bottom) as a function of Bob's reduced state amplitude imbalance $\alpha^2 - \beta^2$. Circles (red) correspond to the original assemblage $\asmbx{\Sigma}{(\alpha)}$ shared as a base for the copies; crosses (blue) are post-selected successfully distilled assemblages, obtained when Bob obtains the outcome $\omega=0$ on his local filter; triangles (orange) correspond to the resulting average assemblages obtained by applying protocol \ref{protocol} to two copies of the original assemblage. Successful distillation can be observed as the values of both measures increase after the process, even for imbalances as high as $0.81$.}
\label{Fig:StDist_ExperimentalResults}
\end{figure}

The subsequent stage of the setup, with Alice's and Bob's devices, is illustrated in Fig.\ \ref{Fig:StDist_Setup} (c). On Alice's side, two black boxes are implemented with half waveplates ($H_{A\,p}$, $H_{A\,s}$), quarter waveplates ($Q_{A\,p}$, $Q_{A\,s}$), a beam displacer ($BD_A$) and a polarizing beam splitter ($PBS_A$). These components allow implementation of a fixed set of projections on both DOF utilized  \cite{Aguilar2014}. Inputs and outputs of the boxes are respectively given by the waveplates' angles and by the photon counts. Conditioning on Bob's state is implemented by coincidence detection.

On Bob's side, local filtering is implemented before photon detection. This is done with a variable mirror (VM) whose reflectance and transmittance depend on its position. The mirror acts only on the lower path on Bob's side, which is tailored to be more intense than the upper path. The VM is set so that transmission of the photon equalizes the amplitudes of both paths, thus implementing $K^{(0)}$  [Eq.\ \eqref{eq:KrausRep_OutSuccess}]. Reflection, on the other hand, corresponds to $K^{(1)}$ and completely destroys steering in the path DOF -- Polarization is then used to prevent weakening of the final correlation.

Photon detectors after the VM register the outcomes, communication to Alice's side is done also through coincidence detection.
Quantum state tomography of all assemblage components $\sigma_{a\vert x}$ is then realized and steering is analyzed on the reconstructed assemblage. Fully characterized components $H_{B,s}$, $Q_{B,s}$, $PBS_{B,s}$, $BD_B$ allow path mode tomography on the transmitted path; waveplates $H_{B,p}$, $Q_{B,p}$, and $PBS_{B,p}$ on the reflected path allow for polarization tomography. 

The results are shown in Fig.\ \ref{Fig:StDist_ExperimentalResults}. Singlet-assemblage fractions for the distilled and original assemblages are shown as function of the population imbalance $\alpha^2 - \beta^2$ of Bob's reduced state. We also show the singlet-assemblage fraction for the case of post-selection, where only the singlet assemblage is kept, given occurrence of a successful local filtering. Distillation with few copies is revealed, considering the intermediate assemblage obtained by combining successful and unsuccessful runs of the protocol using experimentally obtained probabilities for each outcome $\omega$. 

As defined here, the singlet-assemblage fraction is not a steering monotone (meaning that it may increase under 1WLOCCs). Hence, another comparison is made using the steering LHS-robustness, which is a proper steering monotone that measures the amount of unsteerable noise that a given assemblage tolerates before becoming itself unsteerable \cite{PW15, SABGS16, Gallego2015}. This is shown in the bottom part of Fig.\ \ref{Fig:StDist_ExperimentalResults}. The same qualitative behavior can be observed for the robustness, with an increase observed for both the intermediate assemblage and the post-selected assemblage. Both observations then demonstrate the successful experimental distillation of an assemblage with stronger steering than initially.

\emph{Concluding remarks.} -- We have devised a steering distillation protocol inspired on the original local-filtering protocol for entanglement distillation \cite{BennettDist_PRA1996}, but exploiting quantum control only on one party. In other words, our protocol works with local operations assisted by classical communication from the untrusted part to the trusted one, a strict subclass of the LOCCs used in entanglement distillation that meets the natural restrictions of the semi-DI scenario. In contrast, we note that Ref. \cite{Pramanik2019} also studies a steering-filtering protocol but to demonstrate steering superactivation, so it is not only device-dependent (exploiting quantum control at both sides) but also probabilistic. Our protocol concentrates steering deterministically. Moreover, it works both in the limiting case of an infinite number of copies of the initial assemblage and in the non-asymptotic regime. In fact, we have experimentally demonstrated it for 2 copies of an input assemblage, each one encoded in a different degree of freedom (polarization or spatial) of the same twin photon pair. To the best of our knowledge, this is the first experimental demonstration of quantum steering distillation. 

Our results offer a number of exciting open problems. E.g., does the converse of steering distillation, i.e. dilution, exist? If so, are these two processes reversible? Or is there a notion of bound steering analogous to bound entanglement \cite{Horodecki1998}? Finally, in our specific setup the quantum state underlying the assemblage also has its entanglement concentrated by the steering distillation protocol. Is this a generic feature, or can one distill steering without the underlying device-dependent process distilling entanglement?


\begin{acknowledgments}
We thank Rodrigo Gallego for helpful discussions and inspiration. The authors acknowledge financial support from the Brazilian agencies CNPq (PQ grants 311416/2015-2, 304196/2018-5 and INCT-IQ), FAPERJ (PDR10 E-26/202.802/2016, JCN E-26/202.701/2018, E-26/010.002997/2014, E-26/202.7890/2017), CAPES (PROCAD2013), and the Serrapilheira Institute (grant number Serra-1709-17173).
\end{acknowledgments}



%


\onecolumngrid
\appendix

\section{Properties of the Assemblage Fidelity}
\label{app:FidProp}

Here we prove that many properties of the usual fidelity $\mathcal{F}$, defined by its application on two bounded positive semidefinite operators $A$ and $B$ as $\mathcal{F}(A,B) = \Tr[\sqrt{\sqrt{A}\,B\,\sqrt{A}}]$, are carried over to the assemblage fidelity $\afid$, introduced in Eq.\ \eqref{eq:Asmb_Fidelity} and recalled here for convenience:
\begin{equation}
\afid(\asmb{\Sigma},\,\asmb{\Xi}) = \min_{x \in [m]} \sum_{a \in [o]} \mathcal{F}(\sigma_{a|x},\,\xi_{a|x}),
\label{eq:App_AsmbFid_Def}
\end{equation}
where $\asmb{\Sigma}$ and $\asmb{\Xi}$ are generic assemblages with same number of inputs and outputs and with components acting on the same Hilbert space $\mathcal{H}_B$.

A possible motivation for definition \eqref{eq:App_AsmbFid_Def} is as follows: Assume that Alice and Bob share assemblages $\asmb{\Sigma} = \{ \sigma_{a|x} \}_{a,x}$ and $\asmb{\Xi} = \{ \xi_{a|x} \}_{a|x}$ and both parties want to assess if the assemblages are different or not. Let Alice then determine her input $x$ according to a strategy that can be encoded in a distribution $P_X(x)$. If Alice and Bob later combine their results for a series of uses of the black boxes and measurements on Bob's quantum states from both assemblages, they can compute the overall average fidelity between the assemblages as
\begin{align}
\mathcal{G}(\asmb{\Sigma},\,\asmb{\Xi}) 
&\coloneqq \sum_{a \in [o],\,x \in [m]} \mathcal{F}(P_X(x)\,\sigma_{a|x},\,P_X(x)\,\xi_{a|x}) \\ 
&= \sum_{a \in [o],\,x \in [m]} P_X(x)\,\mathcal{F}(\sigma_{a|x},\,\xi_{a|x}).
\label{eq:App_AvgFidelity}
\end{align}
Since their objective is to distinguish the assemblages, Alice uses her ability of determining $P_X$ to ensure that any difference between the assemblages will contribute to lower the value of \eqref{eq:App_AvgFidelity}. For that she considers the worst-case scenario and picks the distribution $P_X$ that minimizes the overall fidelity. The assemblage fidelity is then established as the result of this minimization:
\begin{equation}
\mathcal{F}_{\textrm{A}}(\asmb{\Sigma},\,\asmb{\Xi}) = \min_{P_X}\,\sum_{a \in [o],\,x \in [m]}\,P_X(x)\, \mathcal{F}(\sigma_{a|x},\,\xi_{a|x}),
\label{eq:App_AsmbFid_AltDef}
\end{equation}
which can be proven to be equal to \eqref{eq:App_AsmbFid_Def}: Since any choice of $P_X$ results in a convex combination of the terms $\sum_{a \in [o]} \mathcal{F}(\sigma_{a|x},\,\xi_{a|x})$, the choice that minimizes $\mathcal{G}$ is a deterministic distribution that consistently returns the value of $x$ for which $\sum_{a \in [o]} \mathcal{F}(\sigma_{a|x},\,\xi_{a|x})$ is minimal.

We now proceed to list and prove the properties of $\afid$ as defined in Eq. \eqref{eq:App_AsmbFid_Def}. First, it should be remarked that $\afid$ reduces to the fidelity between classical distributions \cite{NielsenChuang2011} when Bob's states are the same, i.e. when $\sigma_{a|x}/\Tr[\sigma_{a|x}] = \xi_{a|x}/\Tr[\xi_{a|x}],\,\forall a,\,x$ [or simply for all $a$ for the particular value of $x$ that minimizes the expression on the right-hand side (rhs) of Eq. \eqref{eq:App_AsmbFid_Def}]. In this case, $\mathcal{F}(\sigma_{a|x},\xi_{a|x}) = \sqrt{P^{\Sigma}(a|x)\,P^{\Xi}(a|x)}$, where $P^{\Sigma}(a|x) \coloneqq \Tr[\sigma_{a|x}]$ and $P^{\Xi}(a|x) \coloneqq \Tr[\xi_{a|x}]$, and thus
\begin{equation}
\afid(\asmb{\Sigma},\,\asmb{\Xi}) = \min_{x \in [m]}\sum_{a \in [o]} \sqrt{P^{\Sigma}(a|x)\,P^{\Xi}(a|x)}.
\label{eq:App_AsmbFid_EqualState}
\end{equation}
If, on the other hand, Bob's states are different, but the conditional distributions $P^{\Sigma}$ and $P^{\Xi}$ are equal, then the assemblage fidelity is simply the expected value of the fidelity of Bob's states--i.e. using $P(a|x) \coloneqq P^{\Sigma}(a|x) = P^{\Xi}(a|x)$, $\rho^{\Sigma}_{a,x} \coloneqq \sigma_{a|x}/P(a|x)$ and $\rho^{\Xi}_{a,x} \coloneqq \xi_{a|x}/P(a|x)$ (ignoring possible pairs $(a,x)$ where $P(a|x)=0$), then 
\begin{equation}
\afid(\asmb{\Sigma},\,\asmb{\Xi}) = \min_{x \in [m]}\sum_{a \in [o]} P(a|x)\,\mathcal{F}(\rho^{\Sigma}_{a,x},\,\rho^{\Xi}_{a,x}).
\label{eq:App_AsmbFid_EqualProb}
\end{equation}

Another property of $\afid$ is that it is symmetric with respect to the interchange of its arguments, which is clear from its dependence on the usual quantum fidelity $\mathcal{F}$: Since interchanging $\asmb{\Sigma}$ and $\asmb{\Xi}$ in $\afid$ amounts to interchanging $\sigma_{a|x}$ and $\xi_{a|x}$ in each application of $\mathcal{F}$, and since $\mathcal{F}$ is symmetric on its arguments, then 
\begin{equation}
\afid(\asmb{\Sigma},\asmb{\Xi}) = \afid(\asmb{\Xi},\asmb{\Sigma}).
\label{eq:App_AsmbFid_Sym}
\end{equation}

Also, since $\mathcal{F} \geq 0$ for any pair of positive semidefinite operators, the minimization over possible values of $x$ in $\afid$ is then a minimization over different possible combinations of nonnegative terms. Consequently, $\afid \geq 0$. Now, to prove that $\afid(\asmb{\Sigma},\,\asmb{\Xi}) \leq 1$ with equality holding iff $\asmb{\Sigma} = \asmb{\Xi}$, we first note that, since $\mathcal{F}(\rho^{\Sigma}_{a|x},\,\rho^{\Xi}_{a|x}) \leq 1$, then
\begin{equation}
\afid(\asmb{\Sigma},\,\asmb{\Xi}) \leq \min_{x \in [m]}\,\sum_{a \in [o]}\,\sqrt{P^{\Sigma}(a|x)\,P^{\Xi}(a|x)}.
\label{eq:App_AsmbFid_UpperBound_Ineq}
\end{equation}
We recognize on the rhs the classical fidelity between two distributions. Since this fidelity is also not greater than 1, then $\afid(\asmb{\Sigma},\,\asmb{\Xi}) \leq 1$.

Finally, to prove the equivalence $\asmb{\Sigma} = \asmb{\Xi} \iff\afid(\asmb{\Sigma},\,\asmb{\Xi}) = 1$, note that in one way, the implication is straightforward: If $\asmb{\Sigma} = \asmb{\Xi}$, then Eq. \eqref{eq:App_AsmbFid_EqualState} reveals that $\afid(\asmb{\Sigma},\,\asmb{\Xi}) = \min_{x \in [m]}\,\sum_{a \in [o]} P(a|x)$. But, since $P(a|x)$ is normalized, $\afid(\asmb{\Sigma},\,\asmb{\Xi}) = 1$, which proves the claim $\asmb{\Sigma} = \asmb{\Xi} \Rightarrow \afid(\asmb{\Sigma},\,\asmb{\Xi}) = 1$.

In the other direction, if $\afid(\asmb{\Sigma},\,\asmb{\Xi})=1$, from Eq. \eqref{eq:App_AsmbFid_UpperBound_Ineq} we conclude that $1 \leq \sum_{a \in [o]}\,\sqrt{P^{\Sigma}(a|x)\,P^{\Xi}(a|x)},\,\forall x \in [m]$. This can only be true if $P^{\Sigma} = P^{\Xi}$, since the maximal attainable value of these expressions is 1 when $P^{\Sigma}(a|x) = P^{\Xi}(a|x),\,\,\forall a \in [o]$.

With this last result, we can use Eq. \eqref{eq:App_AsmbFid_EqualProb} to write
\begin{equation}
1 = \min_{x \in [m]} \sum_{a \in [o]} P(a|x)\,\mathcal{F}(\rho^{\Sigma}_{a,x},\,\rho^{\Xi}_{a,x}).
\label{eq:App_AsmbFid_EqProof}
\end{equation}
Since each sum is a weighted average of terms upper-bounded by 1, the expression can only be satisfied if $\mathcal{F}(\rho^{\Sigma}_{a,x},\,\rho^{\Xi}_{a,x}) = 1$ for all $(a,x)$ such that $P(a|x) > 0$. This, in turn, implies that $\rho^{\Sigma}_{a,x} = \rho^{\Xi}_{a,x}$ for all such $(a, x)$. Since the assemblage is obtained by combining the probability $P(a|x)$ with Bob's quantum state as $\sigma_{a|x} = P(a|x)\,\rho_{a,x}$, then either $P(a|x) > 0$ and $\sigma_{a|x} = \xi_{a|x}$, or $P(a|x)=0$ and $\sigma_{a|x} = 0 = \xi_{a|x}$. Consequently, for all $a \in [o],\,x \in[m],\, \sigma_{a|x} = \xi_{a|x}$ and, therefore, $\asmb{\Sigma} = \asmb{\Xi}$, which concludes the proof.

\section{Proof of theorem \ref{Thm:SteerDistillation}}
\label{app:Proof}

In what follows, we prove that distillation of quantum steering is possible both in the asymptotic regime of infinite copies and in the finite-copies regime. It should be remarked that our particular solution to this problem allows for perfect distillation of a singlet assemblage even when the initial assemblage is almost unsteerable, given that enough copies are provided. This ability is, however, a feature of the model used for the initial assemblage and is not necessarily a characteristic present in a more general case.

Assume that Alice and Bob share initially $N$ copies of the assemblage $\asmbx{\Sigma}{(\alpha)} \coloneqq \{ \sigma^{(\alpha)}_{a|x} \}_{a \in [2], x \in [2]}$, given by the components
\begin{subequations}
\label{eq:App_AsmbAlpha}
\begin{align}
\sigma^{(\alpha)}_{0|0} = \alpha^2\,|0\rangle\langle 0|,&\quad
\sigma^{(\alpha)}_{1|0} = \beta^2\,|1\rangle\langle 1|,
\label{eq:App_AsmbAlpha_Comp10} \\
\sigma^{(\alpha)}_{0|1} = \frac{1}{2}\,|\alpha_+\rangle\langle \alpha_+|,
&\quad \sigma^{(\alpha)}_{1|1} = \frac{1}{2}\,|\alpha_-\rangle\langle \alpha_-|,
\label{eq:App_AsmbAlpha_Comp11}
\end{align}
\end{subequations}
where $\alpha, \beta$ are real coefficients satisfying $1>\alpha>\beta>0$ and $\alpha^2+\beta^2=1$, and $|\alpha_{\pm}\rangle \coloneqq \alpha|0\rangle \pm \beta|1\rangle$. Following the steps of protocol \ref{protocol}, first Bob applies the POVM $\boldsymbol{M}=\{M^{(\omega)}\}_{\omega\in[2]}$ on the quantum state of the first copy of $\asmbx{\Sigma}{(\alpha)}$. 

We assume that both $M^{(\omega)}$ admit the decomposition $M^{(\omega)} = K^{(\omega)\,\dagger} K^{(\omega)}$, with $K^{(\omega)}$ defined as in Eqs.\ \eqref{eq:KrausRep}. For a given outcome $\omega$ of the POVM, the assemblage is transformed as
\begin{equation}
\sigma^{\prime}_{a|x,\omega} = \frac{K^{(\omega)}\,\sigma^{(\alpha)}_{a|x}\,K^{(\omega)\,\dagger}}{\Tr\left[K^{(\omega)}\,\rho_B\,K^{(\omega)\,\dagger}\right]},
\label{eq:App_KrausMap}
\end{equation}
where $\rho_B$ is the reduced state on Bob's side, $\rho_B = \sum_a \sigma_{a|x}$.

The trace in the denominator of \eqref{eq:App_KrausMap} is the probability of obtaining outcome $\omega$ in each application of $\boldsymbol{M}$. Outcome 0 thus occurs with probability $\Tr\left[K^{(0)} \rho_B K^{(0)\,\dagger} \right] = 2\,\beta^2$. If this outcome is obtained, computation of Eq. \eqref{eq:App_KrausMap} for every component of the assemblage reveals that $\asmbx{\Sigma}{(\alpha)}$ is mapped onto $\asmbx{\Sigma}{\Phi^+}$ [Eq.\ \eqref{eq:SingAsmb}] and the distillation is therefore successful. 
In the case that outcome 1 is obtained, the state on Bob's side is mapped onto $K^{(1)}\rho_{a,x}K^{(1)\,\dagger} \propto |0\rangle\langle 0|$ and the corresponding copy of $\asmbx{\Sigma}{(\alpha)}$ becomes unsteerable. In either case, Bob adds the outcome obtained, $\omega_i$, where $i$ is the index of the current copy, to the string $\boldsymbol{\omega}$ and proceeds to the next copy. 

The previous process of measurement, verification and update of $\boldsymbol{\omega}$ is repeated until Bob reaches the last assemblage. At this point, Bob does not operate on the remaining copy and either adds $\omega_N=1$ to $\boldsymbol{\omega}$ if any previous outcome $\omega_i$ is equal to 0, or adds $\omega_N=0$ otherwise. Bob then sends $\boldsymbol{\omega}$ to Alice and both parties discard all systems marked with $\omega_i=1,\,i=1,...,N$.

Since $\boldsymbol{M}$ is applied independently on each copy, the probability that 
only outcome 1 is obtained for $1 \leq i \leq N-1$ is given by $P_{\textrm{fail}} = \left(\Tr[K^{(1)}\,\rho_B\,K^{(1)\,\dagger}]\right)^{N-1} = (1 - 2\,\beta^2)^{N-1}$. After Bob's message to Alice, the parties then manage to either keep at least one successfully distilled assemblage $\asmbx{\Sigma}{\Phi^+}$ with probability $P_{\textrm{success}} = 1 -~(1-2\,\beta^2)^{N-1}$, or the last copy of $\asmbx{\Sigma}{(\alpha)}$ with probability $P_{\textrm{fail}}$. Consequently, $P_{\textrm{success}} \rightarrow 1$ as $N \rightarrow \infty$ and the protocol ensures distillation of at least one copy of a singlet assemblage in the asymptotic regime. Moreover, since the success or failure of each attempt follows a binomial distribution $P_{\textrm{bin}}(n;\,p) = \binom{N}{n}\,p^n\,(1-p)^{N-n}$ with success probability $p = 2\beta^2$, the rate of extracted singlet assemblages is given by $2\,\beta^2\,(N-1)/N$, which converges to $2\,\beta^2$ as $N \rightarrow \infty$.

For a finite number of copies, a single assemblage can be extracted $\asmbx{\Sigma}{\textrm{dist}}$ as a convex combination of $\asmbx{\Sigma}{\Phi^+}$ and $\asmbx{\Sigma}{(\alpha)}$. The components are given by
\begin{subequations}
\label{eq:App_DistAsmb}
\begin{align}
\sigma^{\textrm{dist}}_{0|0} = \left(\frac{1}{2}\,P_{\textrm{success}} + \alpha^2 P_{\textrm{fail}}\right)|0\rangle\langle 0|,&\quad \sigma^{\textrm{dist}}_{1|0} = \left(\frac{1}{2}\,P_{\textrm{success}} + \beta^2 P_{\textrm{fail}}\right)|1\rangle\langle 1|, 
\label{eq:App_DistAsmb_x0}
\\
\sigma^{\textrm{dist}}_{0|1} = \frac{1}{2}\,\left(P_{\textrm{success}}\,|+\rangle\langle+| + P_{\textrm{fail}}|\alpha^+\rangle\langle \alpha^+|\right),&\quad 
\sigma^{\textrm{dist}}_{1|1} = \frac{1}{2}\,\left(P_{\textrm{success}}\,|-\rangle\langle -| + P_{\textrm{fail}}|\alpha^-\rangle\langle \alpha^-|\right).
\label{eq:App_DistAsmb_x1}
\end{align}
\end{subequations}

Since the singlet assemblage $\asmbx{\Sigma}{\Phi^+}$ is pure, the singlet-assemblage fraction can be rewritten as
\begin{equation}
\singfid(\asmbx{\Sigma}{\textrm{dist}}) = \frac{1}{\sqrt{2}}\,\min_{x \in [2]} \sum_{a \in [2]} \sqrt{\langle \Phi_{a,x}|\sigma^{\textrm{dist}}_{a|x}| \Phi_{a,x}\rangle},
\label{eq:App_ASF_Form2}
\end{equation}
where $\frac12 |\Phi_{a,x}\rangle\langle \Phi_{a,x}| = \sigma^{\Phi^+}_{a|x}$ [Eqs.\ \eqref{eq:SingAsmb}]. Thus, using Eqs. \eqref{eq:App_DistAsmb}, calculating the singlet-assemblage fraction reduces to evaluating the expression
\begin{equation}
\singfid(\asmbx{\Sigma}{\textrm{dist}}) = 
\min 
\left\{ \frac12\left(\sqrt{1 + P_{\textrm{fail}}(\alpha^2 - \beta^2)} + \sqrt{1 - P_{\textrm{fail}}(\alpha^2 - \beta^2)}\right),\, 
\sqrt{1 - \frac12 P_{\textrm{fail}}\,(\alpha - \beta)^2}
\right\}.
\label{eq:App_ASF_Reduced}
\end{equation}

Now, recall that $P_{\textrm{fail}} = (\alpha^2 - \beta^2)^{N-1}$ and let $F_0 \coloneqq \frac12 [\sqrt{1 + (\alpha^2 - \beta^2)^N} + \sqrt{1 - (\alpha^2 - \beta^2)^N}\,]$, and $F_1 \coloneqq \sqrt{1 - \frac12 (\alpha^2-\beta^2)^{N-1}\,(\alpha-\beta)^2}$, such that $\singfid(\asmbx{\Sigma}{\textrm{dist}}) = \min\{F_0,\,F_1\}$. Let also $\Delta \coloneqq \alpha^2 - \beta^2,\, u \coloneqq 2\,F_0^2 - 1,\,v \coloneqq 2\,F_1^2 - 1$ and notice that
\begin{align}
u  &= \sqrt{1 - \Delta^{2N}} 
\label{eq:App_DefU} \\
v  &= 1 - \Delta^{N-1}\,(\alpha-\beta)^2.
\label{eq:App_DefV} \\
\end{align}
Since $\Delta,\,(\alpha-\beta) \leq 1$, both $u$ and $v$ are nonnegative. Consider now $u^2-v^2$, a straightforward algebraic manipulation leads to the expression
\begin{equation}
u^2 - v^2 = 2\,\Delta^{N-1}\,(1-2\alpha\beta)(1-\Delta^{N-1}).
\label{eq:App_diffUV}
\end{equation}
Since each term in the expression is nonnegative, we may conclude that $u^2 \geq v^2$, which implies $u \geq v$, since $u,\,v\geq 0$. Therefore, $F_0 \geq F_1$ for all values of $\alpha$ in the interval $[1/\sqrt{2},\,1]$. This in turn implies that 
\begin{equation}
\singfid(\asmbx{\Sigma}{\textrm{dist}}) = F_1 = \sqrt{1 - \frac12(\alpha^2-\beta^2)^{N-1}(\alpha-\beta)^2},\quad \forall \alpha \in (1/\sqrt{2},\,1),
\end{equation}
which concludes the proof.


%

\end{document}